\documentstyle[preprint,aps]{revtex}
\tightenlines
\begin{document}
\begin{center}
 EFFECT OF SURFACE INDUCED NUCLEATION OF FERROELASTIC DOMAINS ON POLARIZATION SWITCHING 
 IN CONSTRAINED FERROELECTRICS\\
\vskip0.5cm
Rajeev Ahluwalia$^1$ and Wenwu Cao$^2$\\
 $^1$Theoretical Division, Los Alamos National Laboratory, Los Alamos, New Mexico 87545\\
 $^2$Department of Physics and Materials Science, City University of Hong Kong, Kowloon, Hong Kong, China 
\end{center}

\vskip0.5cm
\begin{abstract}
The role of the surface during polarization switching in constrained ferroelectrics is investigated using the 
time-dependent Ginzburg-Landau theory. The model incorporates the elastic and electrostrictive effects in 
the form of a long-range interaction that is obtained by eliminating the strain fields subject to the elastic 
compatibility constraint. A square shaped finite sized constrained ferroelectric system with vanishing surface 
polarization is considered. Computer simulations of the hysteresis reveal that the corners of the constrained 
square act as nucleation sources for $90^{o}$ domain structures. 
 It is also found that no nucleation of $90^{o}$ domain structure takes place in a 
system with semi-infinite (corner free)
geometry. For the corner free case, the polarization switches by $180^{o}$ reorientations and a front of the reverse domains emerges from the surface, eventually sweeping over the entire system. Size dependence of the hysteresis loops is also studied.
\end{abstract}

\eject

\section{Introduction}
The spontaneous polarization in ferroelectrics can be switched by applying a strong electric field 
opposite to the polarization direction \cite{glass}. It is crucial to understand the details of this switching
process due to many applications in memory devices and electromechanical systems. It is
now a well established fact that the switching behavior is strongly influenced by nucleation, growth and motion of domain walls.
Due to the nucleation events, 
the experimentally measured coercive field is much smaller
than the {\it intrinsic} value predicted by the simple 1-D
Landau theory \cite{intr}. In a recent work \cite{prb}, we qualitatively showed that dipolar defects can act as nucleation 
centres for $90^{o}$ domain structures. 
However, for
a constrained system, such as a grain in a ferroelectric ceramic, the surface itself is a heterogeniety that can cause
localized nucleation. Although, there have been some studies investigating the role of a surface on ferroelectric 
phase transitions \cite{su1,su2}, the polarization switching in finite sized systems has not been well studied theoretically.

In this paper, we study the influence of a surface during polarization switching using the 
time dependent Ginzburg-Landau model described in         
\cite{prb}. In particular, we are interested in understanding the role played by the surface in nucleating $90^{o}$ ferroelastic domains during the switching process. Recent experiments on $BaTiO_{3}$ single crystals \cite{bhatta} have demonstrated the importance of $90^o$ domains during polarization reversal. The theory incorporates elastic and electrostrictive effects in the form of an anisotropic 
long-range interaction obtained by eliminating strain fields, subject to the elastic compatibility requirement. 
Instead of using periodic boundary conditions (corresponding to a bulk system), we consider a finite sized 
constrained ferroelectric with the boundary condition of vanishing polarization at the surface. Physically, this 
corresponds to a system where the free charge at the surface is compensated. It is 
believed that this charge compensation occurs at the grain boundaries in ferroelectric ceramics. Due to the above discussed boundary 
conditions, we can neglect the effects of the depolarization field. The nonlocal interaction is represented by the appropriate polarization gradients terms in the free energy expansion.

The organization of this paper is as follows. In section 2, we give details of the time dependent 
Ginzburg-Landau model, starting
from a free-energy functional for the polarization field. Section 3 gives details of our simulations of polarization 
switching for a finite square shaped system. In section 4, we describe simulations of switching behavior in a semi-infinite constrained system. We end the 
paper by a summary and discussion of our results in section 5.
\section{Model}
The 3-D Ginzburg-Landau free energy 
for ferroelectric systems has been given in Ref. \cite{cao}. 
In this work, we restrict to a 2-D system undergoing a square to rectangle 
transition, for which the total free-energy is given by
\begin{equation}
F=\int d\vec{r}  \big[f_{l} + f_{g} + f_{el} + f_{es} + f_{ext} \big]
\end{equation}
 where $f_{l}$ is the local free-energy density given by      
\begin{eqnarray}
f_{l}={{\alpha_{1}}\over{2}} ({P_x}^2 + {P_y}^2)+
{{\alpha_{11}}\over{4}} ({P_x}^4 + {P_y}^4)+
{{\alpha_{12}}\over{2}}{{P_x}^2}{{P_y}^2},
\end{eqnarray}
where $P_x$ and $P_y$ are the components of the polarization vector, $\alpha_{1}$, $\alpha_{11}$ and $\alpha_{12}$ are the free energy parameters.
The term $f_{g}$ represents the gradient energy, 
\begin{eqnarray}
f_{g}={{g_1}\over{2}}\bigg[\bigg({{\partial P_{x}}\over{\partial x}}\bigg)^2
+\bigg({{\partial P_{y}}\over{\partial y}}\bigg)^2 \bigg]+
{{g_2}\over{2}}\bigg[\bigg({{\partial P_{x}}\over{\partial y}}\bigg)^2+
\bigg({{\partial P_{y}}\over{\partial x}}\bigg)^2\bigg]
+
{{g_3}\over{2}}\bigg({{\partial P_{x}}\over{\partial x}}\bigg)
\bigg({{\partial P_{y}}\over{\partial y}}\bigg),
\end{eqnarray}
where $g_1$, $g_2$ and $g_3$ are the gradient parameters.
It has been shown that the gradient energy may be obtained as a local approximation to the nonlocal
interactions \cite{glass}. The term $f_{el}$ represents the usual elastic energy of the square system. We 
consider the bulk strain $\phi_1 =(\eta_{xx} +\eta_{yy})/\sqrt{2}$, 
 deviatoric strain $\phi_2 =(\eta_{xx} -\eta_{yy})/\sqrt{2}$ 
and shear strain $\phi_3 =\eta_{xy} =\eta_{yx}$. Here $\eta_{ij}$ is the linear elastic strain tensor given as
$\eta_{ij}={{1}\over{2}}({{\partial u_{i}}\over{\partial x_{j}}}
+{{\partial u_{j}}\over{\partial x_{i}}})$, where $u_{i}$, $(i=x,y)$ represents the dispacement components.  
The elastic free-energy
can then be written as 
\begin{eqnarray}
f_{el}={{a_1}\over{2}}{\phi_{1}}^2
+{{a_2}\over{2}}{\phi_{2}}^2
+{{a_3}\over{2}}{\phi_{3}}^2,
\end{eqnarray}
where $a_1$, $a_2$ and $a_3$ are constants that can be expressed in terms of linear combinations of elastic constants of the system.
The coupling energy $f_{es}$ may be written as, 
\begin{eqnarray}
f_{es}=-{q_1}{\phi_{1}}({P_{x}}^2 + {P_{y}}^2)
-{q_2}{\phi_{2}}({P_{x}}^2 - {P_{y}}^2)
-{q_3}{\phi_{3}}{P_x}{P_y}.
\end{eqnarray}
Here $q_1$, $q_2$ and $q_3$ are constants that are related to the electrostrictive constants of the system.
The term $f_{ext}$ is given by
$f_{ext}=-{\vec{E}_{ext}}\cdot{\vec{P}}.$
The amplitude of field ${\vec{E}}_{ext}$ is a tunable parameter in our simulations for studying polarization 
switching.  
We assume that the system reaches mechanical equilibrium very fast so that we may integrate out the elastic
fields, subject to the 
elastic compatibility constraint \cite{love}. In terms of the strain components $\phi_1$, $\phi_2$ and
$\phi_3$, the elastic compatibility relation is given as
\begin{equation}
{{\nabla}^2}{\phi_1}-({{\partial^2}\over{\partial x^2}}
-{{\partial^2}\over{\partial y^2}}){\phi_2}-\sqrt{8}
{{\partial^2}\over{ {\partial x}{\partial y} }}{\phi_3}=0.
\end{equation}
To incorporate the elastic compatibility constraint, we consider an effective elastic part of the free energy
$F_{eff}=F_{el}+F_{es} + F_{cons}$, with Lagrangian multiplier $\lambda$, where
\begin{equation}
F_{el}=\int d \vec{k}\bigg\{ {{a_1}\over{2}}{|{\phi_1}(\vec{k})|^{2}} 
+ {{a_2}\over{2}}{|{\phi_2}(\vec{k})|^{2}}
+ {{a_3}\over{2}}{|{\phi_3}(\vec{k})|^{2}}\bigg\}, 
\end{equation}
\begin{eqnarray}
F_{es}=-\int d\vec{k}\bigg\{{q_1}{\Gamma_1}(\vec{k}){\phi_1}(-\vec{k})
+{q_2}{\Gamma_2}(\vec{k}){\phi_2}(-\vec{k})
+{q_3}{\Gamma_3}(\vec{k}){\phi_3}(-\vec{k})\bigg\}
\end{eqnarray}
\begin{eqnarray}
F_{cons}=\int d \vec{k}\bigg \{\lambda(\vec{k})(-{k^2}{\phi_1}(-\vec{k})
+({k_x}^2-{k_y}^2){\phi_2}(-\vec{k})
+{\sqrt{8}}{k_x}{k_y}{\phi_3}(-\vec{k}))
\bigg\}.
\end{eqnarray}
Here $k^2={k_x}^2 +{k_y}^2$ and $\Gamma_1(\vec{k})$
, ${\Gamma_2}(\vec{k})$
 and ${\Gamma_3}(\vec{k})$ are, respectively, the Fourier transforms of 
 ${P_x}^2 +{P_y}^2$, 
 ${P_x}^2 -{P_y}^2$, 
 and ${P_x}{P_y}$. The condition of mechanical equilibrium and the compatibility relation demands 
${{\delta F_{eff}}\over{\delta {\phi_{i}}}}=0$$(i=1,2,3)$ and
${{\delta F_{eff}}\over{\delta {\lambda}}}=0$. 
\begin{eqnarray}
{a_1}{\phi_1}(\vec{k})&=&{q_1}{\Gamma_1}(\vec{k})+{k^2}
\lambda(\vec{k})\nonumber\\
{a_2}{\phi_2}(\vec{k})&=&{q_2}{\Gamma_2}(\vec{k})-
({{k_x}^2-{k_y}^2})
\lambda(\vec{k})\nonumber\\
{a_3}{\phi_3}(\vec{k})&=&{q_3}{\Gamma_3}(\vec{k})-{\sqrt{8}
{k_x}{k_y}}
\lambda(\vec{k}),
\end{eqnarray}
subject to the constraint
\begin{equation}
-{k^2}{\phi_1}(-\vec{k})
+({k_x}^2-{k_y}^2){\phi_2}(-\vec{k})
+{\sqrt{8}}{k_x}{k_y}{\phi_3}(-\vec{k})=0.
\end{equation}
Using the above equations, the multiplier $\lambda(\vec{k})$ is given as 
\begin{equation}
\lambda(-\vec{k})={{{{-k^2 q_1 \Gamma_1(-\vec{k})}/{a_1}}+
{{({k_x}^2-{k_y}^2) q_2 \Gamma_2(-\vec{k})}/{a_2}}
+{{{\sqrt{8}{k_x}{k_y} q_3 \Gamma_3(-\vec{k})}/{a_3}}}}\over{
{ {{k^4}/{a_1}}+ {{({k_x}^2-{k_y}^2)^2}/{a_2}} +
 {8{{k_x}^2}{{k_y}^2}/{a_3}}}}}.
\end{equation}
At this stage we make a simplification by assuming that the polarization electrostrictively couples 
only to the deviatoric strain. This assumption will not effect the qualitative behavior of the 
model in context 
surface effects, which are the main focus of this paper. We should also keep in mind that for the present case of a square to rectangle ferroelectric transition, the deviatoric strains are the dominant mode of deformation. The bulk and shear strains are usually localized at the domain walls for this case. 
Including the coupling with bulk and shear deformations may introduce intermediate 
metastable states with more complex patterns. Exclusion of these couplings makes the computation faster due to faster equilibriation. Finally, we should also mention that the neglect of this coupling
in the free energy does not meen that bulk and shear strains will not be generated in this model at all. 
In fact, in this model, the strains are related to each other through the elastic compatibility relations which leads to the creation of bulk and shear strains, even under the above described approximation.
In the limit of no coupling of polarization
with shear and bulk strains, i.e. $q_{1} \rightarrow 0$
and $q_{3} \rightarrow 0$, using the equilibrium conditions and the lagrange multiplier $\lambda(\vec{k})$, the free energy $F_{eff}$ can be expressed in Fourier space as  
\begin{equation}
F_{eff}= {{{q_2}^2}\over {2{a_2}}} \int d\vec{k}H(\vec{k}) 
|{\Gamma_2}(\vec{k})|^2,
\end{equation} where
$H(\vec{k})= ({ {{h_1}^2(\vec{k})} \over 
{{\alpha}} }+
{ {{h_2}^2(\vec{k})-2{h_2}(\vec{k})} }+
{ {{h_3}^2(\vec{k})} \over {{\beta}} })$.
The quantities 
${h_1}(\vec{k})={k}^2 Q(\vec{k})$, 
${h_2}(\vec{k})=\{1-({k_x}^2-{k_y}^2)Q(\vec{k})\}$
and ${h_3}(\vec{k})=
-{\sqrt{8}}{k_x}{k_y}{Q}(\vec{k})$,
with $Q(\vec{k})$ defined as
\begin{equation}
Q(\vec{k})=
 {{({k_x}^2-{k_y}^2)}\over 
{( {k^4}/{\alpha}+ ({k_x}^2-{k_y}^2)^2 +
 8{{k_x}^2}{{k_y}^2}/{\beta})}}.
\end{equation}
 We have introduced dimensionless constants $\alpha={a_1}/{a_2}$ and 
$\beta={a_3}/{a_2}$. 

The effective interaction derived above is strongly direction dependent and is crucial to describe 
the domain wall orientations. Similar anisotropic interactions have been considered in context of martensitic
 transformations \cite{Kartha,Shenoy}.

As we wish to study heterogeneous nucleation, a dynamical formalism is 
needed to 
describe the non-equilibrium effects associated with domain switching. 
The time dependent Ginzburg-Landau model for the polarization fields is used.
 We introduce rescaled time variables
$t^*=({t}|\alpha_{1}|L)$$(\alpha_{1} < 0)$( $L$ is the kinetic coefficient ), 
space variable $\vec{r^*}= {\vec{r}}/{\theta}$, the wave vector $\vec{k^*}=\vec{k} \theta$
$(\theta=\sqrt{{g_1}/a|\alpha_{1}|})$, where $a$ is a dimensionless constant. 
The polarization field is transformed as
${P_{x}}={P_{R}}u$ and 
${P_{y}}={P_{R}}v$, where  
$P_{R}=\sqrt{|\alpha_{1}|/\alpha_{11}}$ is the remnant polarization 
of the homogeneous state without elastic effects. With this set of parameters, 
the dimensionless
 time dependent Ginzburg-Landau equations are given as
\begin{eqnarray}
{u,}_{t^*}=u-u^{3}-duv^{2}
+a{u,}_{{x^*}{x^*}}
+b{u,}_{{y^*}{y^*}}
+c{v,}_{{x^*}{y^*}}+e_{x}
-{\gamma}u\int d \vec{k^*}H(\vec{k^*})\Gamma(\vec{k^*})exp(-i\vec{k^*}\cdot\vec{r^{*}})\nonumber\\
{v,}_{t^*}=v-v^{3}-dvu^{2}
+a{v,}_{{y^*}{y^*}}
+b{v,}_{{x^*}{x^*}}
+c{u,}_{{x^*}{y^*}}+ e_{y}
+{\gamma}v\int d \vec{k^*}H(\vec{k^*})\Gamma(\vec{k^*})exp(-i\vec{k^*}\cdot\vec{r^{*}}).
\end{eqnarray}
Here, $\vec{e}={\vec{E}_{ext}}/({P_{R}}{|\alpha_1|})$ is the rescaled external electric field. The notation
${u,}_{t}$ represents the time derivative.  
The defined constants are: 
$b=({g_2}/|\alpha_1|{\theta}^2)$, 
$c=({g_3}/|\alpha_1|{\theta}^2)$,   
$d=({\alpha_{12}}/{\alpha_{11}})$ and 
$\gamma=({q_2}^2/{\alpha_{11}}{a_2})$. The quantity 
$\Gamma(\vec{k^*})$ is the Fourier transform of $(u^2-v^2)$.

\section{Simulation of polarization switching for a constrained square shaped 
ferroelectric}
  We now describe the details of our simulations on the switching behavior for 
a finite constrained ferroelectric. The long-range interaction is
conveniently defined in Fourier space, which allows us to use the Fourier Transform method.
This enforces the need to have periodic boundary conditions at the surfaces. To
define a finite system using this scheme, we consider an $88\times 88$ square
region within a full simulation grid of 
size $128 \times 128$. The constraining is implemented by assuming that the polarization 
variables $u(\vec{r^*})$ and $v(\vec{r^*})$ 
vanish at the boundary and outside of the $88 \times 88$ region. This configuration 
represents an electromechanically clamped ferroelectric in which all displacements vanish at the surface.
 Equation (15) is discretized using the Euler scheme on the above described 
grid.  
The space discretization step $\Delta x^*=\Delta y^*=1 $ and the time interval
$\Delta t^*=0.02$. The free energy parameters chosed by us are $d=2$ and
$\gamma=0.05$. The gradient coefficients are chosen as $a=b=c=1$. 
Similarly, we 
choose the elastic parameter ratios as $\alpha=\beta=1$.  
 We first start from a random 
small amplitude initial conditions for the fields $u(\vec{r^*})$ and 
$v(\vec{r^*})$ and an external field $e_{x^*}=10$. We solve equation (15) to
obtain a single domain configuration with the polarization oriented towards the
$+x^*$  gradually decaying at the surface of the square system. Figure 1(a)
shows this configuration corresponding to an electric field value $e_{x^*}=10$. 
In this figure, the arrows represent the local polarization vector 
( we have not shown all the polarization vectors for the sake of 
legibility). Notice the gradual decay of the amplitude of the vectors in the surface 
region. This
configuration is used as the initial state for the simulations of polarization switching. Equations (15) are now solved 
at each time step, starting from the initial 
configuration depicted in figure 1(a) with a time dependent external 
field $e_{x^*}(t^*)$. The time dependence is given as $e_{x^*}=e_0[1-2{t^*}/T]$, where $e_0=10$ is the initial electric 
field value and $T=5000$ is the total number of simulated switching time steps. As the field decreases, the magnitude of the polarization vectors 
decreases uniformly every where in the simulated system at the beginning (keep in mind that the polarization is fixed to zero at the surface
for all times). As the external electric field approaches zero, we observe  
that near two of the corners, the polarization vectors are distorted towards the [11] directions. Figure 1(b) shows the pattern corresponding to $e_{x^*}=0.22$ where we can see the diagonal orientation of the polarization vectors near the corners. This indicates that shear 
strains are generated near the corners. The result is in agreement with recent simulations by Jacobs for a constrained ferroelastic system 
undergoing a square to rectangle transition where it was demonstrated that nondeviatoric strains are generated at the corners 
\cite{jacobs1,jacobs2}.  
These diagonal shear strains at the corners are responsible for nucleation of
$90^{o}$ domain walls, as seen figure 1(c)$(e_{x^*}=0.199)$ and figure 1(d)$(e_{x^*}=0.0)$. 
In figure 1(e) $(e_{x^*}=-0.199)$, we can clearly see   
a well developed twinned 
structure with $90^{o}$ domain walls. As we further decrease the field to $e_{x^*}=-0.38$, domains polarized along 
$[01]$ direction grow at the expense of domains polarized along $[10]$ directions, as can 
be seen by comparing figure 1(e) and figure 1(f). In figure 1(f) $(e_{x^*}=-0.38)$, we 
can also observe that the dipoles in the vicinity of the corners and domain walls have started reorienting towards the $-{x^*}$ direction. 
Thereafter, the reversed domains grow at the expense of domains
polarized along $[01]$, and eventually all the dipoles are oriented along the
$-x^*$ direction. This growth can be clearly seen from figure 1(g) $(e_{x^*}=-0.46)$
and figure 1(h) $(e_{x^*}=-10)$. Figure 1 illustrates the special role played by the corners during switching. The corners act as nucleation centres for $90^o$ domain structures. Unlike the classically held belief of $180^o$ reorientation of polarization, the switching takes
place by sucessive $90^o$ reorientations of the dipoles. Recently, evidence for such non $180^o$ switching has been observed in rhombohedral
single crystals of PZN-PT \cite{caoyin}. The reverse cycle can be simulated by using the configuration in figure 1(h) as the initial condition and increasing the field as $e_{x^*}=e_0[1-2{t^*}/T]$, where $e_0=-10$ and $T=5000$ is the total number of switching time steps. Figure 2 shows the 
hysteresis loops obtained by plotting the average polarization $<u>$ with the field $e_{x^*}$. The solid line corresponds to the situation depicted in figure 1 and the dashed line corresponds to this loop for the case without any surface (periodic boundary conditions). We can observe that for the case of constrained 
square, the simulated hysteresis loop shows a long "tail" region where the forward and reverse cycles curves do not close up. This 
corresponds to the
appearence and disappearence of the $90^o$ domain pattern. In contrast, for the 
periodic boundary condition case, hysteresis loop has no "tail" region as  
there are no nucleation events. We can also observe that the saturation polarization and remnant polarization field are 
lower for the constrained system than that for the case with periodic boundary conditions. However, there is not much difference between the coercive fields of the two cases.
We have simulated the hysteresis 
loops for different system sizes to study the size dependence of the
switching phenomena. In figure 3, we show the simulated hysteresis loops for system sizes indicated at the top of each frame. A 
comparison of the loops for sizes $L=88$ and $L=68$ reveals that the coercive fields and the saturation polarization are almost the same 
for both systems. Also, the $90^o$ domain structure is shorter lived for the smaller system since the "tail" region of the hysteresis loop for the $L=68$ system is smaller than that for the $L=88$ system. For smaller sizes, the 
domain pattern becomes increasingly shortlived during the switching process, as can be concluded from the loop for $L=48$ where the tail has moved further inwards. However, the saturation polarization and the coercive fields still do not show much variation. The tail almost disappears for the $L=20$ hysteresis loop and we see a small bump near the coercive field region. For this size, we can see a reduction in the coercive field as well as the saturation polarization due to the surface effects.
 Below this size, nucleation of $90^o$ domains does not take place. The switching occurs by $180^o$ reorientations that 
preferentially starts at the surface. The hysteresis loops in this regime have shape similar to the periodic boundary condition case. 
It is clear from figure 3 that on further decreasing the size,  
the 
saturation polarization, remnant polarization and coercive field decrease rapidly. We can see that the hysteresis is very small for
$L=6$ which is the smallest size simulated. For the present set of parameter values, we do not observe a ferroelectric to paraelectric 
transition for $L=6$. Such a transition was observed for a simulation with much higher value of the gradient coefficients.
 The behavior
observed in figure 3 can be understood in terms of size dependence of 
ferroelectric behavior. It is experimentally and theoretically established that
the spontaneous polarization progressively decreases on decreasing size for a
system in which the polarization is constrained to decay at the surface 
\cite{su1,su2,payne,rjap}. Thus, the remnant polarization, saturation 
polarization and coercive field decrease on decreasing the system size.
The anisotropic
nonlocal interaction responsible for the nucleation of $90^o$ domain
structures also decreases as the size is decreased. Thus for small sizes, the
hysteresis loops are free of $90^o$ nucleation. A similar transition from multidomain 
to single domain states on decreasing the size has been discussed in 
earlier works \cite{rjap,arlt}. The main effect of size reduction comes from the surface layer under the vanishing polarization boundary condition. The transition temperature is effectively reduced and the energy barrier between the low temperatures states is also lowered.

\section{Simulation of polarization switching for a semi infinite constrained ferroelectric}
In this section, we describe switching behavior for a 
semi-infinite constrained ferroelectric system. Here we use periodic boundary
conditions along the $y$ direction and vanishing polarization conditions at 
the surface for $x$ direction. Figure 4 shows the dynamics of polarization 
switching for the semi-infinite case. Except the boundary conditions, all other
parameter values, including the time dependence of the switching field are the same 
as that for the case of the constrained square described in section 3. Figure 4(a) shows the fully polarized
 single domain initial configuration for
$e_{x^*}=10$. In figure 4(b), we can see that a single domain persists even when the applied field is 
lowered to $e_x=-0.319$, although the polarization magnitude has decreased by a very small amount.
As the field is further decreased to 
$e_{x^*} \sim -0.36$, 
 nucleation of the reversed polarization
domains from the two surfaces takes place, as depicted in figure 4(c). The reversed domains grow larger and eventually into a single domain crystal with reversed polarization. 
 This reversed domain growth is clear from figure 4(d) ($e_{x^*}=-0.399$) and figure 4(e) ($e_{x^*}=-0.44$). Figure 4(f)
shows the final state corresponding to $e_{x^*}=-10$. 
 Figure 4 illustrates the fact that for the corner free semi-infinite geometry, the reversal occurs only by $180^o$ reorientations. The hysteresis loop corresponding to the situation shown in figure 4 can be computed. In figure 5, we plot the hysteresis loop for the case shown in figure 4 along with the loop
for the surface free case with the periodic boundary conditions. Unlike the 
case shown in figure 2, the hysteresis loops for the surface free case and the 
semi-infinite case have the same shape. However, the saturation polarization, 
the remnant polarization and the coercive field are slightly smaller for the semi-infinite case due to the suppressed transition at the surface. We have also 
simulated the size dependence for this case and found that the hysteresis loops become progressively narrower with decreasing size without any shape change. This is due to the fact that there are no $90^o$ domains and
the long-range elastic interaction does not play any role in this case.

\section{Summary and discussion}
We have simulated the switching behavior in constrained ferroelectrics with the boundary condition of vanishing polarization at
the surface. A square shaped system is considered along with a 
system with
semi-infinite geometry. We find that the corners of the square shaped system nucleate $90^o$ domains and the switching is 
dominated by the motion of $90^o$ domain walls. In contrast, for the semi-infinite geometry, there is no nucleation of $90^o$ domains and the
switching proceeds by $180^o$ reversal. Clearly, the corners in the constrained square play a crucial role during the switching process.
We believe that the anisotropy of the long-range elastic interaction interacts with the corners to produce inhomogenieties that
result in $90^o$ nucleation.
Our results suggest that the switching behavior and the hysteresis loops in real systems may depend on the shape of the grain and its boundary conditions. 

We have also simulated the size dependence of hysteresis loops. The decrease of the saturation polarization, 
remnant polarization and the coercive field with decreasing size is common to both configurations. This can be
understood in terms of the supression of the transition in an electromechanically constrained system as the system size approaches the correlation length of order parameter. The free energy for a constrained system is a function of the system size. For the constrained square case, the polarization reversal process is dominated by 
intermediate $90^o$ domain pattern changes for larger systems and becomes totally via $180^o$ reorientations for size less than a critical dimension. This transition is due to the weakening of the long-range elastic interactions with decreasing system size.

Finally, one should note that the behavior of decreasing coercive field with decreasing size indicated by the simulation results does not apply to thin films \cite{intr}, but only to clamped grains in ferroelectric polycrystals.  
For thin films, not all the dimensions are reduced simultaneously, which is different than our condition here. In addition, the misfit strain with the substrate is important, which has to be incorporated in the theory \cite{pertsev}.
\section*{Acknowledgement}
This research is sponsored by the Hong Kong Research Grant Council under grant number CityU 1056/02P.

Figure captions:

Figure. 1: Pattern evolution for the simulated switching in a constrained square. The snapshots correspond to 
(a) $e_{x^*}=10$; 
(b) $e_{x^*}=0.22$; 
(c) $e_{x^*}=0.199$; 
(d) $e_{x^*}=0$; 
(e) $e_{x^*}=-0.199$; 
(f) $e_{x^*}=-0.38$; 
(g) $e_{x^*}=-0.46$; 
(h) $e_{x^*}=-10$. 

Figure. 2: Plots of $<u>$ vs $e_{x}$ showing the simulated hysteresis loops for the constrained square
(solid line) and the periodic boundary condition case (dotted line). 

Figure. 3: Simulated hysteresis loops for different sizes of the constrained square.

Figure. 4: Pattern evolution for the simulated switching in a semi-infinite constrained system. The snapshots correspond to 
(a) $e_{x^*}=10$; 
(b) $e_{x^*}=-0.319$; 
(c) $e_{x^*}=-0.36$; 
(d) $e_{x^*}=-0.399$; 
(e) $e_{x^*}=-0.44$; 
(f) $e_{x^*}=-10$. 

Figure. 5: Plots of $<u>$ vs $e_{x}$ showing the simulated hysteresis loops for the semi-infinite constrained system
(solid line) and the periodic boundary condition case (dotted line). 


\begin{thebibliography}{40}             
\bibitem{glass} M. E. Lines and A. M. Glass,{\it Principles and Applications of Ferroelectrics and 
Related Materials} (Clarendon, Oxford, 1979). 
\bibitem{intr} Stephen Ducharme, V. M. Fridkin, A. V. Bune, S. P. Palto, L. M. Blinov, N. N. Petukhova, and
 S. G. Yudin, Phys. Rev. Lett. {\bf 84}, 175(2000).
\bibitem{prb} R. Ahluwalia and W. Cao, Phys. Rev. B {\bf 63}, 012103 (2001).
\bibitem{bhatta} E. Burscu, G. Ravichandran and K. Bhattacharya, Appl. Phys. Lett. {\bf 77}, 1698(2000).
\bibitem{su1} W. Y. Shih, W. H. Shih and I. A. Aksay, Phys. Rev. B {\bf 50}, 15575(1994).
\bibitem{su2} Y. G. Wang, W. L. Zhong and P. L. Zhang, Phys. Rev. B {\bf 51}, 5311(1995).
\bibitem{cao} Wenwu Cao and L. E. Cross, Phys. Rev. B {\bf 44}, 5 (1991).
\bibitem{love} E. A. H. Love, { \it A Treatise on the Mathematical Theory of Elasticity} (Dover, New York, 1944),
p. 49.
\bibitem{Kartha} S. Kartha, J. A. Krumhansl, J. P. Sethna and L. K. Wickham, Phys. Rev. B,
 {\bf 52}, 803 (1995).
\bibitem{Shenoy} S. R. Shenoy, T. Lookman, A. Saxena and A. R. Bishop, Phys. Rev. B {\bf 60}, R12537 (1999).
\bibitem{jacobs1} A. E. Jacobs, Phys. Rev. B {\bf 52}, 6327(1995) 
\bibitem{jacobs2} A. E. Jacobs, Phys. Rev. B {\bf 61}, 6587(2000) 
\bibitem{caoyin} Jianhua Yin and Wenwu Cao, Appl. Phys. Lett., {\bf 79}, 4556(2001).
\bibitem{payne} M. H. Frey and D. A. Payne, Phys. Rev. B {\bf 54}, 3158(1996).
\bibitem{rjap} R. Ahluwalia and W. Cao, J. Appl. Phys. {\bf 89}, 8105(2001)
\bibitem{arlt} G. Arlt, Ferroelectrics {\bf 104}, 217 (1990).
\bibitem{pertsev} N. A. Pertsev and V. G. Koukhar, Phys. Rev. Lett. {\bf 84}, 3722(2000).
\end{thebibliography}
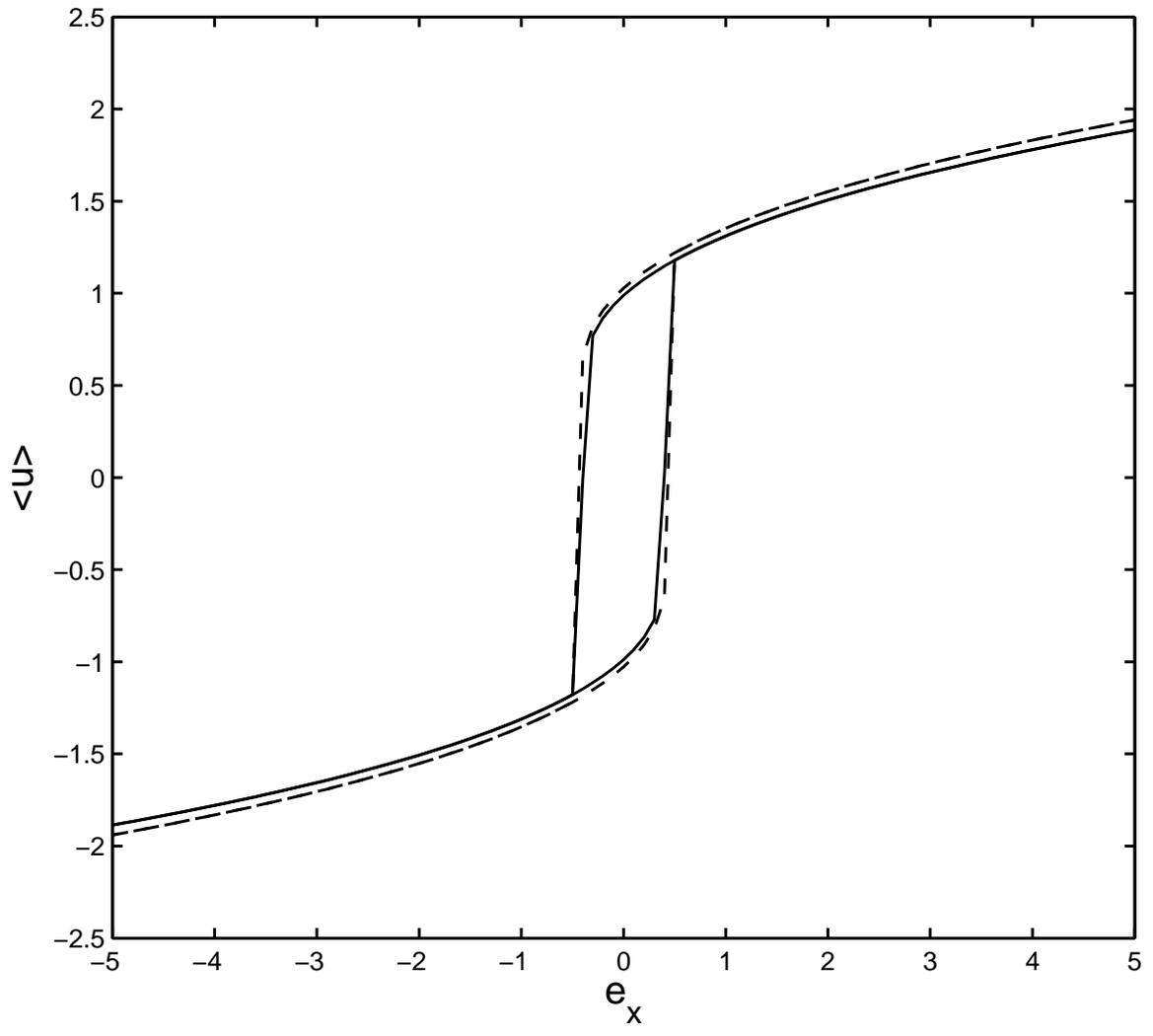
\end{document}